\documentclass[aps,pra,twocolumn,superscriptaddress,showpacs]{revtex4-1}

\usepackage{amsmath}
\usepackage{graphicx}
\usepackage{amssymb}

\newcommand{\ham}[1]{\hat{\mathcal{H}}_{#1}}
\newcommand{\op}[1]{\hat{#1}}
\def\tr{ {\rm{Tr }}}

\begin{document}

\title{Simulation of noise-assisted transport via optical cavity networks}

\author{Filippo Caruso}
\email{filippo.caruso@uni-ulm.de}
\affiliation{Institut f\"{u}r Theoretische Physik, Universit\"{a}t Ulm, Albert-Einstein-Allee 11, 89069 Ulm, Germany}
\author{Nicol\`{o} Spagnolo}
\affiliation{Dipartimento di Fisica, Sapienza Universit\`{a}  di Roma, piazzale Aldo Moro 5, 00185 Roma, Italy}
\affiliation{Consorzio Nazionale Interuniversitario per le Scienze Fisiche della Materia, piazzale Aldo Moro 5, 00185 Roma, Italy}
\author{Chiara Vitelli}
\affiliation{Dipartimento di Fisica, Sapienza Universit\`{a}  di Roma, piazzale Aldo Moro 5, 00185 Roma, Italy}
\affiliation{Consorzio Nazionale Interuniversitario per le Scienze Fisiche della Materia, piazzale Aldo Moro 5, 00185 Roma, Italy}
\author{Fabio Sciarrino}
\email{fabio.sciarrino@uniroma1.it}
\affiliation{Dipartimento di Fisica, Sapienza Universit\`{a}  di Roma, piazzale Aldo Moro 5, 00185 Roma, Italy}
\affiliation{Istituto Nazionale di Ottica, largo Fermi 6, 50125 Firenze, Italy}
\author{Martin B. Plenio}
\affiliation{Institut f\"{u}r Theoretische Physik, Universit\"{a}t Ulm, Albert-Einstein-Allee 11, 89069 Ulm, Germany}

\begin{abstract}
Recently, the presence of noise has been found to play a key role in assisting the transport of energy and information in complex quantum networks and
even in biomolecular systems. Here we propose an experimentally realizable optical network scheme for the demonstration of the basic mechanisms underlying noise-assisted transport. The proposed system
consists of a network of coupled quantum optical cavities, injected with a single photon, whose transmission efficiency can be measured.
Introducing dephasing in the photon path this system exhibits a characteristic enhancement of the transport efficiency that can be observed with presently available technology.
\end{abstract}


\maketitle
\section{Introduction}

The presence of noise in quantum transmission networks is generally considered to be deleterious
for the efficient transfer of energy or classical/quantum information encoded in quantum states.
Quantum networks, used for the transmission, are unavoidably interacting with an
external noisy environment and this interaction significantly affects the quantum coherence of the system
evolution. It is indeed commonly accepted that the presence of decoherence \cite{Zure03} is responsible for the undesired and uncontrolled
transfer of information from the system to the environment, which in turn reduces the coherence in
quantum systems.
However, recently noise has been found to play a positive role in creating quantum
coherence and entanglement \cite{Plen02,Hart06}. Motivated by fascinating experiments showing
the presence of quantum beating in photosynthetic systems \cite{Lee07,Prok02,Enge07}, subsequent theoretical work pointed to the idea that the remarkable efficiency of the excitation energy transfer in light harvesting complexes during photosynthesis benefits from the presence of environmental noise \cite{Mohs08,Plen08}. Indeed, the intricate interplay between dephasing and
quantum coherence and also the entanglement behaviour during the noise-assisted transport dynamics have
been elucidated in more details in Refs. \cite{Rebentrost,Caru09,Caru10,Chin10}. Perhaps even more surprisingly,
the dephasing was recently found to assist the transfer of classical and quantum information in communication complex quantum networks \cite{Caru10a}.
\begin{figure}[t!]
    \centering
        \includegraphics[width=0.49\textwidth]{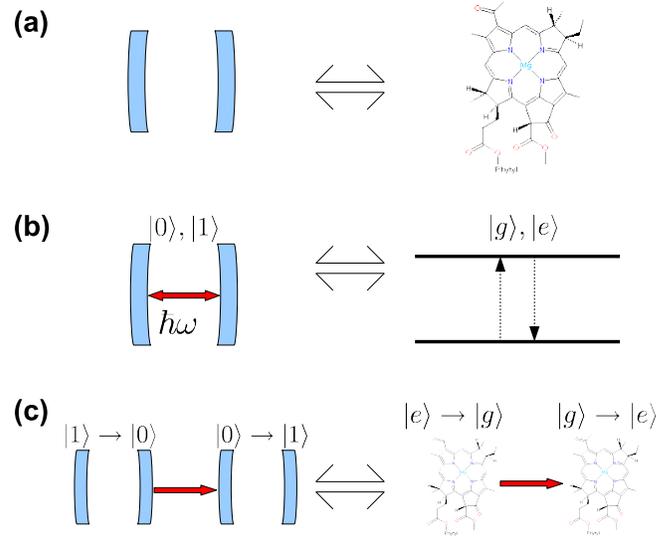}
    \caption{Analogy between a network of coupled optical cavities and a physical network
    where the excitation is carried by electrons, such as light-harvesting complexes.
    (a) The single site of the network is analogous to a single optical
    cavity. (b) The electronic excitation is represented by
    the presence of a single photon in the corresponding optical cavity. (c) The transfer of the excitation
    between two interacting sites is analogous to the transfer of the single-photon between two
    adjacent coupled cavities.}
    \label{fig:comparison_cavity_FMO}
\end{figure}

Recently, quantum optical systems have been exploited as a promising platform to simulate quantum processes \cite{Hart06a,Angelakis,Greentree}.
For example, several implementations of systems simulating quantum random walks have been reported with
linear optical resonators \cite{Bouw99,Knig03}, linear optical elements \cite{Do05,White}, fiber network \cite{Schr10}
and optical waveguides \cite{Pere08,Chri03,OBri09,Peru10}. Motivated by these results, here we propose a ``quantum optical scheme'' to investigate the
noise-assisted excitation transfer process through a set of coupled
optical cavities. We discuss a four-site optical network and derive the set of relevant parameters
that rule the time evolution of the system. A detailed numerical simulation of this
dynamics, when one cavity is injected by a single-photon, is performed employing realistic experimental parameters, showing that
the presence of a suitable dephasing process in each site of the network allows for a characteristic increase of the excitation transfer efficiency.
Furthermore, we consider several aspects such as phase-stabilization of the cavities and the implementation of the dephasing
that are necessary to observe a clear enhancement of the photon transfer rate from one cavity to an external detector mimicking the so-called
reaction center of the light-harvesting complexes. Finally, we investigate how entanglement degrades during the time evolution of the optical network.

The paper is organized as follows. In Sec. \ref{sec:model} we define the model that describes the dynamics of the four-site optical network
analysed in this paper, including the master equation for the two relevant noise processes. Then in Sec. \ref{sec:parameters}
we perform a detailed derivation of a realistic set of parameters for the system. In Sec. \ref{sec:simulation} we report the results of
a numerical simulation of the dynamics of the network. Finally, the conclusions and final remarks are presented in Sec. \ref{sec:conclusions}.
\begin{widetext}

\begin{figure}[hb!]
        \includegraphics[width=.9\textwidth]{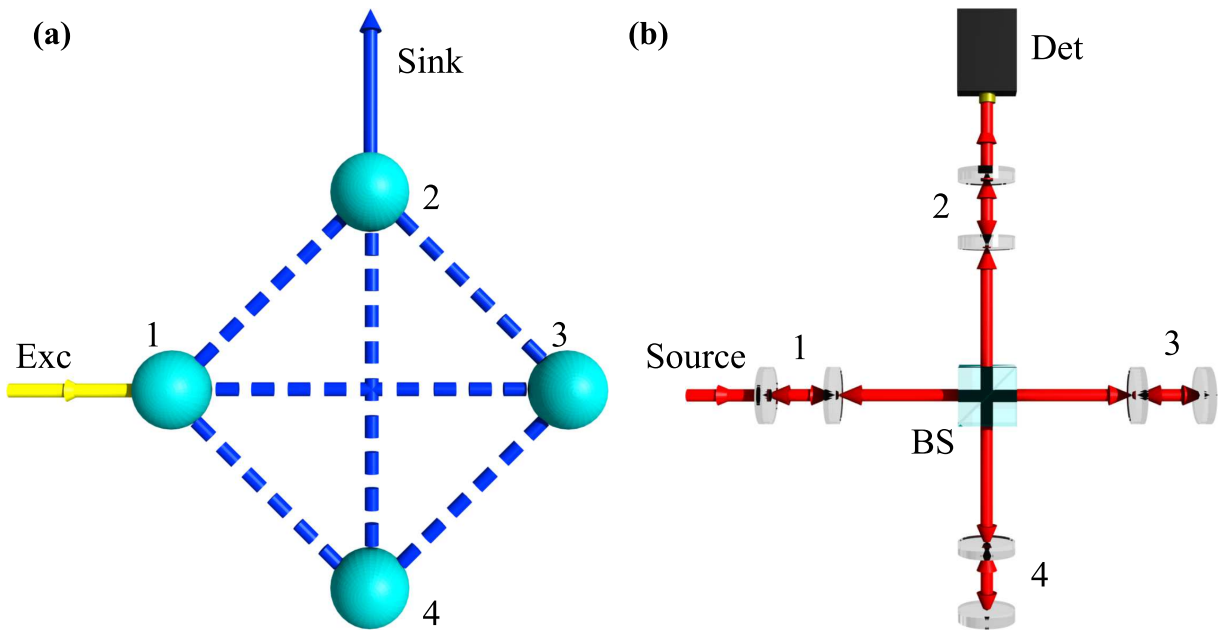}
    \caption{Quantum simulation of noise–assisted excitation
transfer through a network of coupled optical cavities. (a)
Simplified scheme of a four-site fully-connected network. The
excitation is injected in site $1$, and exits from the network by
coupling of site $2$ with the output sink. (b) Equivalent network of
four coupled cavities. The excitation is given by a single-photon
pulse injected in cavity $1$. The right output mode of cavity $2$ is the
output sink channel for the excitation. }
    \label{fig2}
\end{figure}

\end{widetext}
\section{Model of the network}
\label{sec:model}
In this section we describe in details the model underlying the
dynamics of the proposed network of optical cavities -- a schematic view of this system in relation to
the light-harvesting complexes is shown in Fig.\ref{fig:comparison_cavity_FMO}.
Starting from the Hamiltonian describing non-interacting cavities, one has
\begin{equation}
\ham{cav} = \sum_{i} \hbar \omega \op{a}_{i}^{\dag} \op{a}_{i} \; ,
\end{equation}
with $\hat{a}_{i}$ and $\hat{a}_{i}^{\dag}$ being the usual bosonic
field operators, which annihilate and create a photon in cavity $i$,
and $\omega$ being the resonance frequency, which we assume for simplicity to be equal for all cavities.
The transfer of photons between the optical cavities is described by the following Hamiltonian term:
\begin{equation}
\ham{int} = \sum_{(i,j)} \hbar g_{ij} \left( \op{a}_{i}^{\dag} \op{a}_{j} + \op{a}_{i} \op{a}_{j}^{\dag}
\right) \; ,
\end{equation}
where the sum on $(i,j)$ extends over all the connected cavities, and $g_{ij}$ are the coupling
constants.
Moreover, we assume that this dynamics is subject simultaneously
to two distinct noise processes acting on each optical cavity $i$:
\begin{itemize}
\item
a dissipation process that leads to photon loss with rate $\Gamma_i$, described by the following Lindblad superoperator
\begin{eqnarray}
\mathcal{L}_{diss}(\op{\rho}) &=& \sum_{i} \Gamma_{i} \left( - \left\{
\op{a}^{\dag}_{i} \op{a}_{i}, \op{\rho} \right\} +2 \op{a}_{i}
\op{\rho} \op{a}_{i}^{\dag} \right)\; ,
\end{eqnarray}
\item
a pure dephasing process that randomizes the photon phase with rate $\gamma_i$ given by a Lindbladian term which has the form
\begin{eqnarray}
\mathcal{L}_{deph}(\op{\rho}) &=& \sum_{i} \gamma_{i} \left( - \left\{ \op{a}_{i}^{\dag} \op{a}_{i}
\op{a}_{i}^{\dag} \op{a}_{i}, \op{\rho} \right\} + 2 \op{a}_{i}^{\dag} \op{a}_{i} \op{\rho}
\op{a}_{i}^{\dag} \op{a}_{i} \right) \; . \; \ \; \ \;
\end{eqnarray}
\end{itemize}

Besides, the total transfer of the single photon is measured in terms of photons detected on the right-hand side output of the cavity $2$, which represents the so-called
sink or reaction center of the biological systems (here, a single-photon detector), described by the Lindblad operator
\begin{eqnarray}
        {\cal L}_{Det}(\rho) &=& \Gamma_{Det} [2 \op{a}_{Det}^{\dag} \op{a}_{2}
        \rho \op{a}_{2}^{\dag}  \op{a}_{Det} - \{\op{a}_{2}^{\dag}  \op{a}_{Det} \op{a}_{Det}^{\dag} \op{a}_{2},\rho\} ] \; \ \ \ \
\end{eqnarray}
where $\op{a}_{Det}^{\dag}$ describes the effective photon creation operator in the detector and $\Gamma_{Det}$ is
the rate at which the photon irreversibly gets the detector on the right side of the optical cavity $2$ -- see Figs. \ref{fig2} and \ref{fig4}. Hence, the photon transfer efficiency is measured by
the quantity
\begin{equation}
p_{sink}(t) = 2\Gamma_{Det}\int_{0}^t \tr[\rho(t') \op{a}_{2}^{\dag} \op{a}_{2}] \mathrm{d}t' \; .
\end{equation}
In the following numerical simulation, we will assume that there is a single photon initially in the cavity $1$. Notice that,
since our scheme does not involve any non-linear process, a single photon experiment repeated many times exhibits
the same statistics obtained with an injected coherent state \cite{Amse09}.
\section{Parameters of the network}
\label{sec:parameters}
In this section we discuss the experimental details of the optical cavity network setup
sketched in Fig. \ref{fig2}, in order to simulate the mechanisms underlying the noise-assisted
transport phenomena. The excitation of the network, i.e. a single photon at wavelength
$\lambda = 800$ nm, is generated through a heralded single photon source, based on the spontaneous
parametric down-conversion process. We consider the case of a network of $d_{k} = 1$ cm long
cavities. The distance between each cavity and the central beamsplitter (BS), chosen
with transmittivity $\eta = 0.5$, is taken to be $l_{k} = 20$ cm. More specifically, cavity $1$
is chosen with mirror reflectivities $R_{1}^{in} = 0.9$ (for the internal mirror pointing
towards the BS) and $R_{1}^{out}=0.99$ (for the external mirror),
cavity $2$  with $R_{2}^{in}=R_{2}^{out}=0.9$, while cavities $3$ and $4$
with $R_{j}^{in}=0.9$ and $R_{j}^{out}=0.999$. The loss parameter of each cavity $j$ is given by
\begin{equation}
\xi_{j}=\sqrt{R_{j}^{in} R_{j}^{out}e^{-2 d_{j} \alpha_{j}}} \; ,
\end{equation}
with $\alpha_{j} = 0.35$ m$^{-1}$,
while the average number of round trips for a photon in the cavity is given by $m_{j} = (1-\xi_j)^{-1}$.
The parameters adopted for the numerical simulation are respectively
the set of \emph{coupling coefficients} between the cavities of the network, the \emph{dissipation rate}
and the \emph{transmission rate} from site $2$ to the output mode $\mathbf{k}_{det}$ (i.e., the detector).
\subsection{Dissipation rate}
\begin{figure}[t!]
    \centering
        \includegraphics[width=0.48\textwidth]{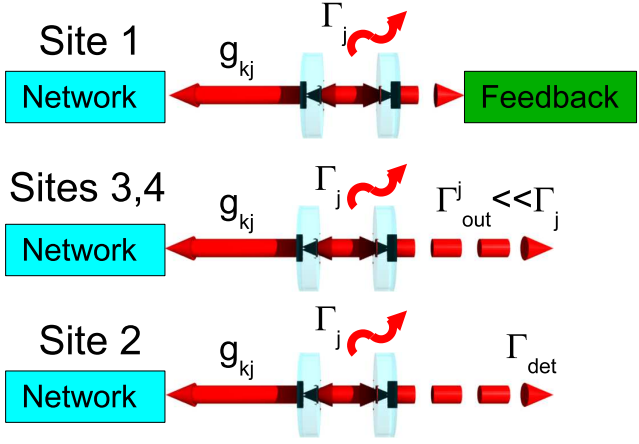}
    \caption{Scheme of all coupling, dissipation and dephasing processes of the network. All sites are coupled with the network with the $g_{kj}$ couplings,
    and undergo both a dephasing process, with rate $\Gamma$, and internal losses, with rate $\Gamma_{j}$. Site 1: external losses are reduced with a feedback system.
    Site 2: external coupling with the sink. Sites 3,4: external losses are negligible with respect to internal losses $\Gamma_{out}^{j} \ll \Gamma_{j}$.}
    \label{fig3}
\end{figure}

The dissipation rate in each cavity $j$ can be evaluated by considering the amount of losses
in $m_{j}$ round trips, i.e. the average flight time of the photon in the cavity. Such parameter
can be evaluated according to the following expression:
\begin{equation}
\label{eq:dissipation_int}
\Gamma_{j} \sim \frac{D}{m_{j} t_{j}} \sum_{i=0}^{m_{j}} \xi_{j}^{2i} = D \frac{1-\xi_{j}^{2(m_{j}+1)}}{(1 - \xi^{2}_{j}) m_{j} t_{j}} \; ,
\end{equation}
where $\xi^{2}_{j}$ represents the fraction of optical power which remains confined in the cavity after each round trip, $t_{j}$
is the photon flight time in one round trip, $m_{j}$ is the average number of round trips for the photon in the cavity $j$,
and $D$ is the dissipation in one round trip only due
to diffraction or coupling with other optical modes, i.e. $D = 1 - e^{-2 \alpha_{j} d_{j}}$.
Eq. (\ref{eq:dissipation_int})
evaluates the fraction of optical power lost in $m_{j}$ round trips, divided by the average flight time $t_{cav} = m_{j} t_{j}=2 d m_{j}/c$ with $c$ being the speed of light. In our setup, one has $\Gamma_{j} \simeq 50 MHz$ and  $D \sim 0.007$.
We consider negligible the losses between the cavities and the beamplitter due to the adoption of anti-reflection coating optics.
Notice that the average flight time of the photon in the cavity, i.e. $t_{cav}$, is of the order of $\mathrm{ns}$
and defines the time scale of the process. Hence, the corresponding linewidth of
the cavity alone, evaluated from the spectral properties of the intracavity field, is of $\sim
2$ GHz. The linewidth of the injected photon must be much smaller than this value, hence narrow band parametric down conversion source,
such as the one obtained through periodically-poled nonlinear crystals,
is necessary for an efficient cavity-photon coupling.
Let us note that the presence of external mirrors with reflectivities $R_{1,2,4}^{out} < 1$
introduces additive channels $\Gamma_{out}$ for losing the photon out of the network in spatial modes $\mathbf{k}_{1,2,4}^{out}$ -- see Fig. \ref{fig3}.
The dissipation rate due to such process is given by the fraction of optical power lost through the external mirror in time $t_{cav} = m_{j} t_{j}$,
and can be evaluated as:
\begin{equation}
\label{eq:dissipation_external}
\Gamma_{out}^{j} \sim  \frac{(1-R_{j}^{out})}{m_{j} t_{j}} \sum_{i=0}^{m_{j}} (R_{j}^{out})^{i} = \frac{1-(R_{j}^{out})^{m_{j}+1}}{m_{j} t_{j}} \; .
\end{equation}
For cavities 3 and 4, such additive dissipation channel is of the order of $\Gamma^{3,4}_{out} \sim 10$ MHz, thus being negligible
with respect to the dissipation due to intracavity losses. For cavity 1, we can introduce a feedback system to discard those events which correspond to losing the photon through this
channel. This system is shown in Fig. \ref{fig4} and exploits the polarization state of the photon by inserting a Faraday rotator (FR). More specifically, the photon with $\vert H \rangle$
polarization state, after the polarization beam-splitter (PBS), is rotated by the $\lambda/2$ waveplate and by the Faraday rotator in the polarization state $\vert V \rangle$. When
the photon exits the network from the external mirror of site $1$, the propagation through the $\lambda/2$ waveplate and the Faraday rotator in the opposite direction maintains its polarization
state $\vert V \rangle$ unaltered. Finally, the photon is reflected by the polarization beam-splitter and then detected.
This allows to discard those events when the detector clicks, thus allowing to reduce the effective dissipation term $\Gamma_{out}^{1}$. We notice that
recent papers reported the realization of high detection efficiency ($\sim 75\%$) silicon avalanche photodiodes and superconducting transition edge detectors, with the
perspective of reaching a value of $\eta_{det} \sim 90\%$ \cite{Thom10,Tsuj10}. The adoption of these devices would reduce the effective dissipation $\Gamma^{1}_{out}$ due to the external mirror by a factor $0.2$, and hence in our case from $\sim 100 MHz$ without post-selection to $\Gamma_{out}^{1} \sim 20 MHz$.
\subsection{Coupling rate}
The cross-coupling coefficients have been evaluated following the theory of Marcuse \cite{Marc85,Marc86}.
The time evolution of the intra-cavity field amplitude is described by the following set of first order differential equations:
\begin{equation}
\label{eq:diff_set_couplings}
\frac{d A^{\nu}_{j}}{dt} = \imath (\Omega^{\nu}_{j} - \Omega) A^{\nu}_{j} + \imath \sum_{\sigma} s_{j}^{\nu \sigma} A^{\sigma}_{j} +
\imath \sum_{k\neq j} \sum_{\sigma} g^{\nu \sigma}_{jk} A_{k}^{\sigma} \; ,
\end{equation}
where $j$ is the cavity index, $(\nu, \sigma)$ are the mode indexes, $s_{j}^{\nu \sigma}$ and $g_{kj}^{\nu \sigma}$ are respectively
the self-coupling and cross-coupling coefficients, $\Omega_{j}^{\nu}$ is the optical frequency corresponding to
the eigenvalue of the propagation equation for the optical mode $\nu$
in cavity $j$, $\Omega$ is a reference frequency corresponding to
the center of the spectrum for $\Omega_{j}^{\nu}$, and $A_{j}^{\nu}$ are the field amplitudes.
In our case, we consider only a single-mode of the electromagnetic field and the mode indexes $(\nu, \sigma)$ can
be neglected. An approximate expression for the evaluation of the cross-coupling coefficients is derived in Ref. \cite{Marc85} and reduces here to the expression:
\begin{equation}
g_{kj} \simeq \frac{1}{2 \imath} \left( \frac{v_{k} v_{j}}{d_{k} d_{j}} \right)^{1/2} t_{kj} \; ,
\end{equation}
where $v_{j}$ are the intracavity group velocities, $d_{j}$ are the cavity lengths and
$t_{kj}$ is the amplitude transmission coefficient between the fields $A_{j}$ and $A_{k}$ in the two sites.
The amplitude transmission coefficient can be directly evaluated by analyzing the fields in the classical limit.
The calculation of this parameter can be divided in three intermediate steps. Particularly, in the following we specify these calculations
for our setup in Fig. \ref{fig2}.

\begin{enumerate}

\item
\emph{Output field from cavity $k$}. As a first step, we evaluate the ratio between the intra-cavity field and the field at the output
of cavity $k$ and it is given by:
\begin{equation}
\frac{A_{out}^{(k)}}{A_{cav}^{(k)}} = \sqrt{(1-R_k^{in})} e^{- \imath \delta \phi_{k}} e^{- \alpha_{k} d_k}
\end{equation}
where $\delta \phi_{k}$ is the phase term due to propagation inside the cavity $k$.

\item
\emph{Inter-cavity field at the input face of cavity $j$.} The field at the input face of cavity $j$ can be
evaluated as the coherent superposition of all possible paths of the output field from site $k$, i.e. $A_{out}^{(k)}$.
Such quantity strongly depends on interference effects between all possible paths that the photons can take in the network.
We restrict our treatment only to the first order path. The ratio between the intercavity field and field at the output of cavity $k$, i.e.
$I_{kj} = \frac{A_{inter}^{(k\rightarrow j)}}{A_{out}^{(k)}}$, has the following form:
\begin{equation}
I_{kj} = \imath^{n_{r}} \frac{1}{\sqrt{2}} e^{\imath(\phi_{k} + \phi_{j})} K_{kj}
\end{equation}
where $n_{r}$ is the number of times the photon has been reflected by the beam-splitter, $\phi_{k} = 2 \pi \frac{l_{k}}{c} \nu_{L}$ is the phase-shift due to spatial propagation between the cavity $k$ and the BS with $\nu_{L}$ being the field optical frequency and $l_k$ being the distance between the cavity $k$ and the central BS, and the form of $K_{kj}$ is
different whether the cavities $k$ and $j$ are directly linked by the beam-splitter or not, i.e.
\begin{equation}
K_{kj} = \left\{ \begin{array}{ll} 1 & \mathrm{for \ direct \ link} \\ \frac{1}{\sqrt{2}}  (\sqrt{R^{in}_{q}} e^{\imath 2 \phi_{q}}+ \sqrt{R^{in}_{p}}
e^{\imath 2 \phi_{p}}) & \mathrm{for \ indirect \ link}\end{array} \right.
\end{equation}
where $p$ and $q$ are cavity indexes satisfying $k \neq j \neq p \neq q$.
In order to generalize the expressions of $I_{kj}$ and $K_{kj}$ to the case of a BS with transmittivity $\eta \neq 0.5$, each factor $\frac{1}{\sqrt{2}}$ has to be replaced with $\sqrt{\eta}$ or $\sqrt{1-\eta}$ whether the photon has been reflected or transmitted by the beam-splitter.

\item
\emph{Intra-cavity field in cavity $j$}. The intra-cavity field in site $j$ is related to the intra-network field
at its input face according to:
\begin{equation}
\frac{E_{cav}^{(j)}}{E_{intra}^{(k \rightarrow j)}} = \frac{\sqrt{(1-R^{in}_{j})}}{1-m_{j}} \; ,
\end{equation}
hence, the coupling coefficients can be finally written as:
\begin{equation}
\begin{aligned}
g_{kj} &= \frac{I_{kj}}{2\imath} \left( \frac{v_{k} v_{j}}{d_k d_j} \right)^{1/2} \frac{\sqrt{(1-R^{in}_{k}) (1-R^{in}_{j})}
e^{- \imath \delta \phi_{k}}}{1-m_{j}} e^{- \alpha_{k} d_k} \;
\end{aligned}
\end{equation}

\end{enumerate}

Following these calculations, the absolute values of the coupling rates
are found to be $g_{12}=4.3$ GHz, $g_{13}=5.7$ GHz, $g_{14}=7.6$ GHz, $g_{23}=6.1$ GHz, $g_{24}=4.5$ GHz and $g_{34}=5.9$ GHz.
To take into account the fluctuations in
the coupling coefficients (induced by the phase fluctuations) in between
experiments, we analyze also the case in which there is a
static disorder of $\sim 20\%$ in the coupling rates.
\subsection{Transmission rate}
Finally, the transmission rate from cavity $2$ to the output mode $\mathbf{k}_{out}$ is evaluated as above for $\Gamma_{j}$, by
considering the amount of field which is transmitted through the mirror $R_{2}^{out}$ in the output mode in $m_2$ round trips. In other words, the rate at which the photon
is transferred to the detector can be evaluated with the same expression for the external mirror dissipation rate of Eq.(\ref{eq:dissipation_external}).
The numerical evaluation of this parameter gives:
\begin{equation}
\Gamma_{Det} = \frac{1-(R_{2}^{out})^{m_{2}+1}}{m_{2} 2d/c} \sim 1 \ GHz \; .
\end{equation}
\begin{widetext}

\begin{figure}[hb!]
        \includegraphics[width=0.9\textwidth]{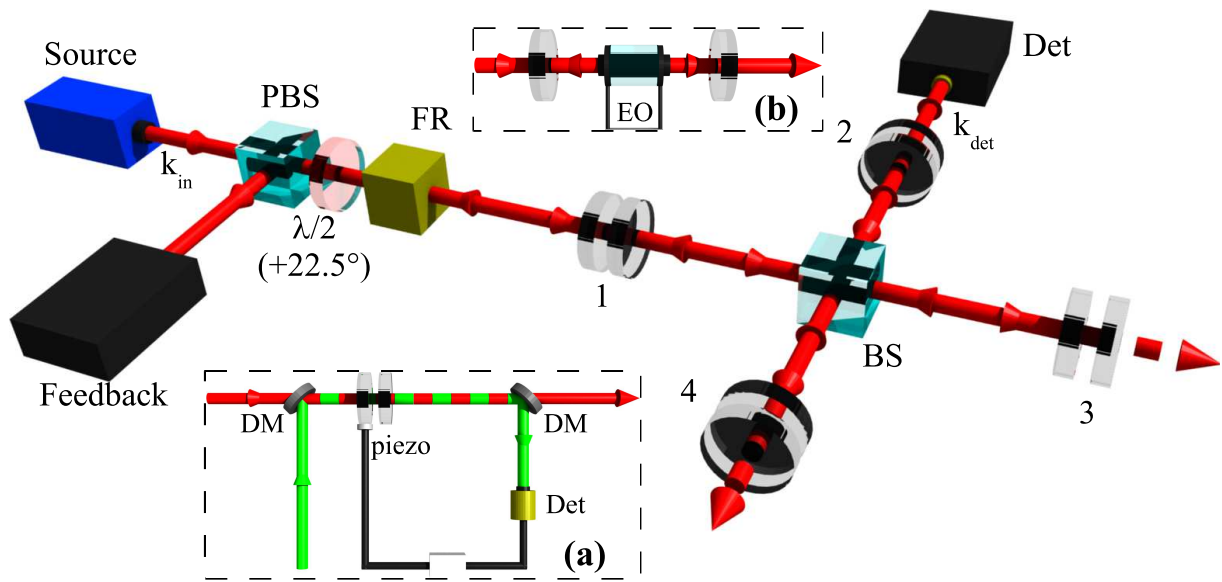}
    \caption{Experimental setup for the four-cavities optical network. The single-photon in the
input mode $\mathbf{k}_{in}$
is injected into the network in site $1$. The successful transfer of the
excitation in the network
is given by the detection of the single-photon on mode
$\mathbf{k}_{det}$. The coupling coefficients
between cavities can be changed by varying the transmittivity and the
reflectivity of the beamsplitter
(BS). The feedback system to reduce the dissipation rate
$\Gamma^{1}_{out}$ from site $1$
exploits the polarization degree of freedom of the photon as described
in the text. [Inset (a)] Sketch of an active phase stabilization apparatus.
An auxiliary laser is injected and extracted in the network by two dichroic mirrors (DM).
The measurement on the auxiliary laser is exploited to drive a piezoelectric system
which stabilize the cavity lenght. [Inset (b)] Introduction
of dephasing rate by modulation of the index of refraction through an electro-optical crystal.}
    \label{fig4}
\end{figure}

\end{widetext}
\subsection{Experimental tasks}
The two main challenges for the proposed experimental
realization, reported in Fig. \ref{fig4}, regard the phase stability of the
cavities and the implementation of a suitable device to introduce the necessary amount of
dephasing rate. An accurate control on the cavity length is necessary in order to maintain the
cavities at resonance with the photon wavelength and to keep the coupling rates constant. This
phase stabilization can be achieved by an active feedback system working on an auxiliary laser
superimposed with the single photon with a dichroic mirror, and then measured after its passage
through the cavity [Fig. \ref{fig4}, inset (a)].

The parameters of the network here adopted correspond to a low finesse optical cavity.
Hence, a length stability of the order of few nanometers is necessary, which is a
requirement fully achievable with the current technology.
The dephasing rate inside the cavity can be introduced by acting on the beam path, through the
propagation factor $e^{ikz}$. This can be done by modifying the propagation length $z$ or by varing
the wavevector $k$, for instance by acting on the index of refraction inside the cavity $n$.
The phase modulation can then be achieved by different methods, depending on the desired modulation
rate. For a dephasing rate of the order of $1$ GHz, as the one exploited in the numerical
simulation below, an electro-optic or acusto-optic modulator can be inserted in the cavity and exploited to
shift the frequency of light [Fig. \ref{fig4}, inset (b)].
\section{Results of the numerical simulation}
\label{sec:simulation}
\begin{figure}[t!]
    \centering
        \includegraphics[width=0.5\textwidth]{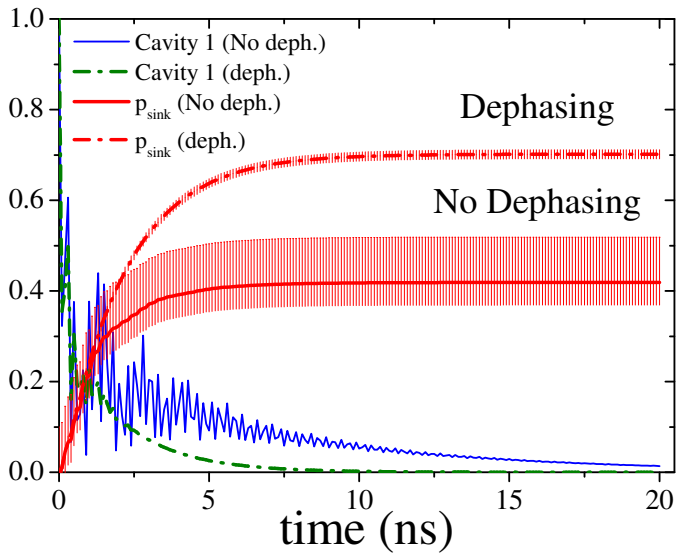}
    \caption{Site $1$ population behaviour and transfer efficiency vs. time (in $\mathrm{ns}$) for the optical cavity network for the noiseless (continuous line) and noise (dashed line) case. The photon transmission is significantly enhanced by the presence of dephasing. The error bars are calculated considering a $20\%$ static disorder ($10^3$ samples).}
    \label{fig5}
\end{figure}
\begin{figure}[t!]
    \centering
        \includegraphics[width=0.5\textwidth]{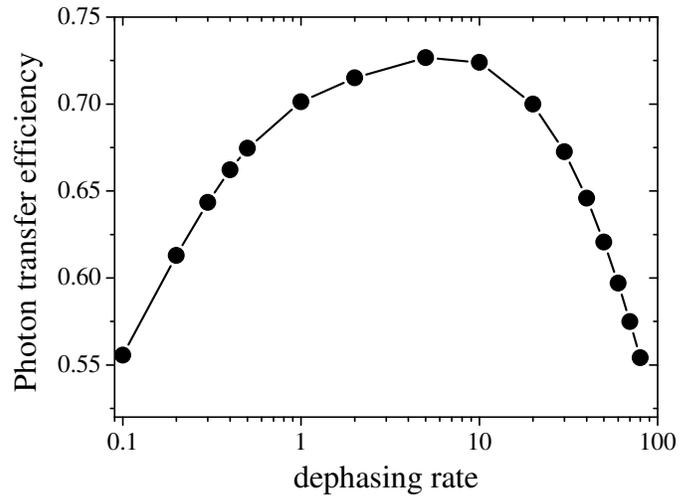}
    \caption{Dependence of $p_{sink}$ at a fixed time $t=20 \ \mathrm{ns}$ as a function of the dephasing rate $\gamma$. These results show the remarkable robustness of this process, supporting the possibility of being experimentally observed.}
    \label{fig6}
\end{figure}

In this section we investigate the dynamics of the optical cavity network, by using our model with the coupling rates $g_{ij}$ as estimated above, the dissipation rate equal to $\Gamma_{j} = 70$ MHz for all the sites by taking into account the different dissipative processes, and the transfer rate to the detector being $\Gamma_{Det} = 1$ GHz. In particular, in Fig. \ref{fig5} we show the cavity photonic population as a function of time in both cases of local dephasing (with rate $1$ GHz) and no dephasing.
The introduction of dephasing reduces the destructive interference between all the possible photon pathways and increases remarkably the overall transfer efficiency from about $40\%$
to more than $70\%$. In absence of dephasing, the photon is trapped in some superposition (\textit{dark}) states which are not coupled to the site $2$ and this explains the lower transfer efficiency - see Ref. \cite{Caru09} for more details. Moreover, we consider also the case in which the coupling rates suffer a static disorder of $20\%$ and also the dependence of the transfer efficiency as a function of the dephasing rate -- see Fig. \ref{fig6}. These results further supports the fact that the noise-assisted transport could be experimentally observed by the present optical setup.

Finally, to quantify the entanglement dynamics, we study logarithmic negativity \cite{Plen07}, i.e. $E(A|B)=\log_2||\rho^{\Gamma_A}||_1$,
measuring the entanglement across a bipartition $A|B$ of a composite system, where $\Gamma_A$ is the partial transpose operation of the density operator $\rho$ with respect to the subsystem $A$ and $|| \cdot ||_1$ denotes the trace norm. In Refs. \cite{Caru09,Caru10}, in the context of the light-harvesting systems, it was found that the increase in the transfer efficiency is not strictly related to the presence of entanglement between the sites of the network and a similar behaviour has been numerically found in this cavity dynamics.
\begin{figure}[h!]
\centerline{\includegraphics[width=.47\textwidth]{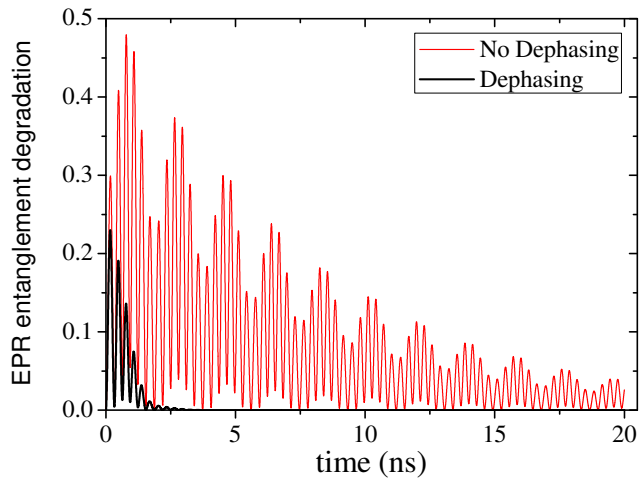}}
\caption{Degradation of entanglement between two photons (initially in a maximally entangled state), in which only one photon is sent to the network through the cavity $1$. It is measured in terms of log-negativity between the ancilla photon and the photon leaving from the cavity $2$ (no detector).}
\label{fig7}
\end{figure}
However, to explore some more the capabilities of this optical cavity network as a conduit for not just energy (or classical information) but
quantum information, we show in Fig.\ref{fig7} how another form of entanglement (in this context more relevant, as measureable more directly) degrades through it.
To that end, we introduce an ancillary photon, which initially shares
a maximally entangled state with the single photon injected in the optical cavity $1$, i.e. in the EPR state $1/\sqrt{2} (|0\rangle_{anc}|0\rangle_{1}+|1\rangle_{anc}|1\rangle_{1})$, with $|0\rangle$ and $|1\rangle$ representing, respectively, the absence and presence of a photon \cite{Lomb02,Giac02}. As the system evolves, the entanglement between the ancillary photon and the photon leaving out from the cavity $2$ oscillates in time and is almost vanishing in presence of dephasing.
\section{Conclusions and Outlook}
\label{sec:conclusions}
We proposed a quantum optical network, based on a set of coupled cavities,
in order to investigate the effects of noise in the excitation transfer. A detailed numerical
simulation for experimentally realistic values shows the presence of a substantial enhancement in
the photon transport efficiency when dephasing noise is introduced in the cavity.
As a final remark, we note that a similar scheme can also be implemented by exploiting a network of atoms
in suitable ion traps \cite{Brow10}. The results reported here may open interesting perspectives for a deeper
investigation of the fundamental mechanisms that underly the very high efficient excitation transfer in
light harvesting complexes and for possible applications in solar cell technology.
Finally, this type of experiments can also trigger significant activity
on two different areas, namely the modeling of complex environments via controlled interactions
and the development of noise--assisted protocols for quantum communication \cite{Caru10a}.
\acknowledgements
This work was supported by the projects AST 2009, HYTEQ (FIRB-MIUR), CORNER, Q-ESSENCE, and the Humboldt Foundation.
F.C. was also supported by a Marie Curie Intra
European Fellowship within the 7th European Community
Framework Programme.
%
%
%

\end{document}